\documentclass[aps,showpacs,prl,twocolumn,superscriptaddress]{revtex4}
\usepackage[pdftex]{graphicx}

\begin{document}
\date{\today}
\title{Long range magnetic ordering in Na$_2$IrO$_3$}

\author{X. Liu}
\affiliation{Condensed Matter Physics and Materials Science Department, Brookhaven National Laboratory, Upton, NY 11973, USA\\}
\author{T. Berlijn}
\affiliation{Condensed Matter Physics and Materials Science Department, Brookhaven National Laboratory, Upton, NY 11973, USA\\}
\affiliation{Physics Department, State University of New York, Stony Brook, New York 11790, USA}
\author{W.-G. Yin}
\affiliation{Condensed Matter Physics and Materials Science Department, Brookhaven National Laboratory, Upton, NY 11973, USA\\}
\author{W. Ku}
\affiliation{Condensed Matter Physics and Materials Science Department, Brookhaven National Laboratory, Upton, NY 11973, USA\\}
\affiliation{Physics Department, State University of New York, Stony Brook, New York 11790, USA}
\author{A. Tsvelik}
\affiliation{Condensed Matter Physics and Materials Science Department, Brookhaven National Laboratory, Upton, NY 11973, USA\\}
\affiliation{Physics Department, State University of New York, Stony Brook, New York 11790, USA}
\author{Young-June Kim}
\affiliation{Department of Physics, University of Toronto, Toronto, Ontario, Canada M5S 1A7\\}
\author{H. Gretarsson}
\affiliation{Department of Physics, University of Toronto, Toronto, Ontario, Canada M5S 1A7\\}
\author{Yogesh Singh}
\affiliation{I. Physikalisches Institut, Georg-August-Universit{\"a}t G{\"o}ttingen, D-37077 G{\"o}ttingen, Germany\\}
\affiliation{Indian Institute of Science Education and Research Mohali, MGSIPAP Complex, Sector 26, Chandigarh 160019, India}
\author{P. Gegenwart}
\affiliation{I. Physikalisches Institut, Georg-August-Universit{\"a}t G{\"o}ttingen, D-37077 G{\"o}ttingen, Germany\\}
\author{J. P. Hill}
\affiliation{Condensed Matter Physics and Materials Science Department, Brookhaven National Laboratory, Upton, NY 11973, USA\\}

\begin{abstract}

We report a combined experimental and theoretical investigation of the magnetic structure of the honeycomb lattice magnet Na$_2$IrO$_3$, a strong candidate for a realization of a gapless spin-liquid. Using resonant x-ray magnetic scattering at the Ir L$_3$-edge, we find 3D long range antiferromagnetic order below T$_N$=13.3 K. From the azimuthal dependence of the magnetic Bragg peak, the ordered moment is determined to be predominantly along the {\it a}-axis. Combining the experimental data with first principles calculations, we propose that the most likely spin structure is a novel ``zig-zag'' structure.

\end{abstract}


\maketitle

Recently, Ir-based transition-metal oxides have attracted significant attention\cite{Mott,Jackeli,pyrochlore1, honeycomb1,pyrochlore2, hyperkagome2Y, hyperkagome2L, hyperkagome2M, hyperkagome2Yi, honeycomb2}, owing to a combination of the strong relativistic spin-orbital coupling (SOC) of the Ir $5d$ electrons and the non-trivial topology of underlying lattice in many of the structural families. These latter include the pyrochlore A$_2$Ir$_2$O$_7$, the hyperkagome structures of A$_4$Ir$_3$O$_8$ and the honeycomb lattices of A$_2$IrO$_3$,  where A is an alkaline metal or lanthanide. These offer the possibility for exotic electronic behavior. Indeed, the strong SOC in the iridates has been shown to lead to novel Mott-insulating states even in the presence of only weak correlations\cite{Mott}. The non-trivial underlying lattice geometries enhance the significance of this coupling because the spin and orbital components of the wavefunction then become entangled, and this can lead to various kinds of strongly anisotropic magnetic interactions, depending on the topology of the underlying lattice \cite{Jackeli}. Thus these materials provide an array of interesting physics, with exotic states such as topological insulators \cite{pyrochlore1, honeycomb1} and spin-liquid states \cite{pyrochlore2, hyperkagome2Y, hyperkagome2L, hyperkagome2M, hyperkagome2Yi, honeycomb2} proposed as possible ground states.

Among the iridates, A$_2$IrO$_3$ is particularly interesting. Here Ir$^{4+}$ ions reside in oxygen octahedra \cite{Felner, Yogesh}, and are suggested to be magnetic with an effective moment, $J_{eff} = 1/2$ \cite{Felner, Yogesh, honeycomb1}. This combination of $J_{eff} = 1/2$ and the underlying honeycomb lattice makes A$_2$IrO$_3$ a promising candidate for the long-sought realization of the exactly solvable Kitaev model\cite{Kitaev1}. This model describes spin S=1/2 magnets with extremely anisotropic exchange on a honeycomb lattice. Its ground state is a gapless spin liquid, which has attracted much attention for its potential in quantum computing applications due to its insensitivity to local perturbations.

Unfortunately, experimental studies of A$_2$IrO$_3$ have been rather limited to date. For Na$_2$IrO$_3$, the only experimental information is the very recent work of Singh {\it et al}. who carried out susceptibility measurements and suggested that Na$_2$IrO$_3$ is antiferromagnetically(AF) ordered at low temperature\cite{Yogesh}, apparently deviating from Kitaev model predictions . This raises the important question of why does the Kitaev model break down in this system? To address this issue, an essential first step is to understand the nature of the low temperature ordered state, including the magnetic structure, the ordered moment direction and dimensionality of the ordering.

Here we report x-ray resonant magnetic scattering studies of Na$_2$IrO$_3$, performed at the Ir L$_3$ edge. Our results explicitly show that Na$_2$IrO$_3$ has a long-range antiferromagnetically ordered ground state below T$_N$=13.3 K and that the ordering is three dimensional(3D). From the azimuthal dependence of the magnetic peaks, the ordered magnetic moment is determined to be mainly along the crystallographic {\it a} direction. Based on our observations, two magnetic ordering structures (see Fig.3) are found to be possible candidates: ``zig-zag'' and ``stripy''. Combining the experimental data with a set of constrained first principles calculations, we propose the zig-zag phase as the most likely ground state.

Na$_2$IrO$_3$ crystals were synthesized by a solid-state synthesis method. The synthesis and bulk property characterization are reported in Ref.\cite{Yogesh}. The x-ray resonant magnetic scattering experiments were carried out at beamline X22C at the National Synchrotron Light Source, Brookhaven National Laboratory. The single crystals used in our experiment were plate-like with typical dimension of $\sim$1x1x0.1 mm$^3$. A Ge(111) analyzer was used to improve the resolution and to suppress the inelastic background due to fluorescence. Except for the temperature dependent work, all the data were collected at 1.58 K.

\begin{figure}[ht]
\includegraphics{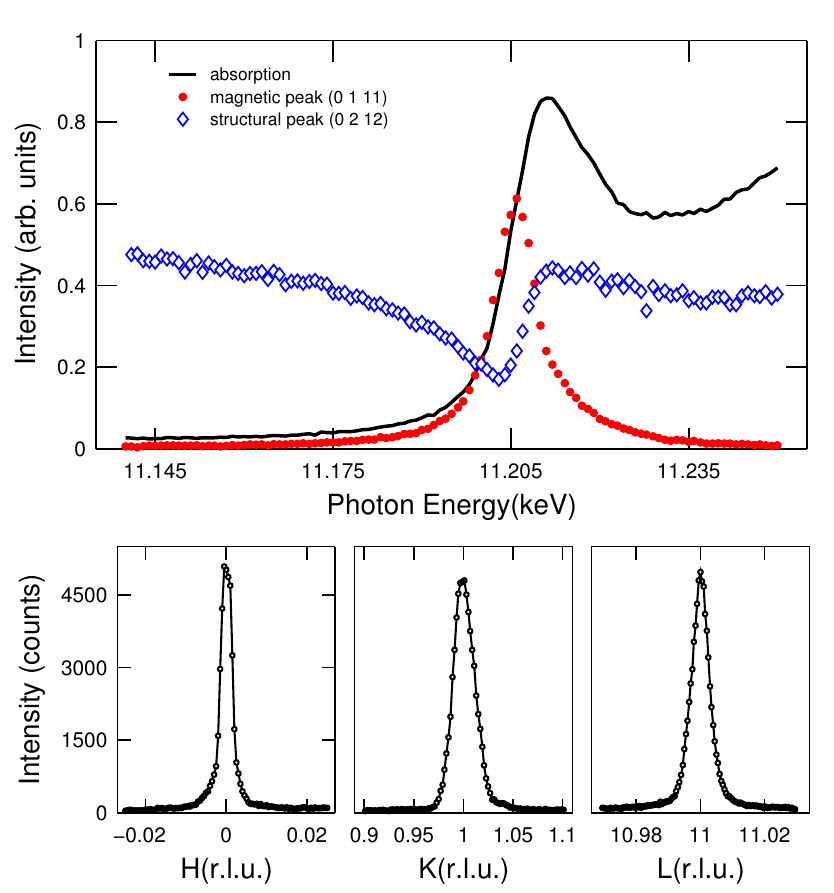}
\caption{\label{fig1}
(Color online). Top panel: Incident photon energy dependence of the (0 1 11) (circles), and the (0 2 12) (diamonds) Bragg peaks on tuning through the Ir L$_3$ edge. The solid line is the absoption spectra of the L$_3$ edge as measured by monitoring the fluorescence signal. Bottom panels: H, K and L scans across the (0 1 11) peak.}
\end{figure}

At low temperatures, well-defined peaks were observed at (0 $\pm$1 L) Bragg points with L=7, 9 and 11 with the incident x-ray energy tuned to the Ir L$_3$ edge. For Na$_2$IrO$_3$ which has a C2/{\it c} structure\cite{Yogesh}, these are structurally forbidden Bragg points. In Fig. 1, we show the energy dependence of one of these peaks, the (0 1 11), near the Ir L$_3$ edge, together with that of the structurally allowed Bragg peak (0 2 12). In contrast to the (0 2 12), the (0 1 11) peak shows a large enhancement on tuning through the Ir L$_3$ edge, as determined by the fluorescence signal (also plotted). In principle, this resonant behavior could be due to either orbital ordering or magnetic ordering. To distinguish the two is non-trivial, especially when there is strong SOC. Based on the fact that the position and shape of this resonance is similar to that previously observed in Sr$_2$IrO$_4$ \cite{Kim} and that the temperature dependence of the ordering is consistent with the susceptibility measurements\cite{Yogesh}, we conclude that the most likely explanation for this peak is that it is magnetic in origin. That is the (0 1 L=odd) reflections demonstrate the presence of long range antiferromagnetic order. 

Scans through the (0 1 11) peak along the three high-symmetry directions are shown in the bottom panels of Fig. 1. The peaks are at, or close to, the resolution limit in all three directions, with the resolution determined by similar scans performed through the nearby (0 2 12) allowed Bragg peak. This indicates that the magnetic ordering is 3D in nature and is long ranged. From the half-width-at-half-maximum (HWHM) of these scans, we place lower limits on the respective magnetic correlation lengths, defined as $(\frac{1}{HWHM})$, to be about $\xi_{mag} \geq$ 600\AA, {120\AA} and {500\AA} along the crystallographic {\it a}, {\it b} and {\it c} directions, respectively. We emphasize that these values largely reflect the instrumental resolution and that the resolution along the K direction is the broadest.

In order to determine the AF ordered moment direction, the azimuthal dependence of the (0 1 11) peak was measured by rotating the sample about the (0 1 11) direction. In our experimental setup, the incident beam is $\sigma$ polarized, that is the electric field is perpendicular to the scattering plane. In this geometry, the dipole resonant magnetic x-ray scattering cross-section varies as ${\hat{k}_f} \cdot {\hat{M}}$, where ${\hat{k}_f}$ is the direction of the scattered beam and ${\hat{M}}$ is the magnetic moment direction\cite{John1996}. In Fig. 2, the intensity of the (0 1 11) magnetic peak is plotted as a function of $\alpha$, where $\alpha$ is the angle between ${\hat{k}_f}$ and the {\it a}-axis. The $\alpha$ dependence of the intensity follows $\cos^2\alpha$ (dashed line) reasonably well, suggesting that the magnetic moment is predominantly along the {\it a}-direction. The slight deviations from this curve may indicate that perhaps the magnetic moment also has some small component in other directions.

\begin{figure}[ht]
\includegraphics{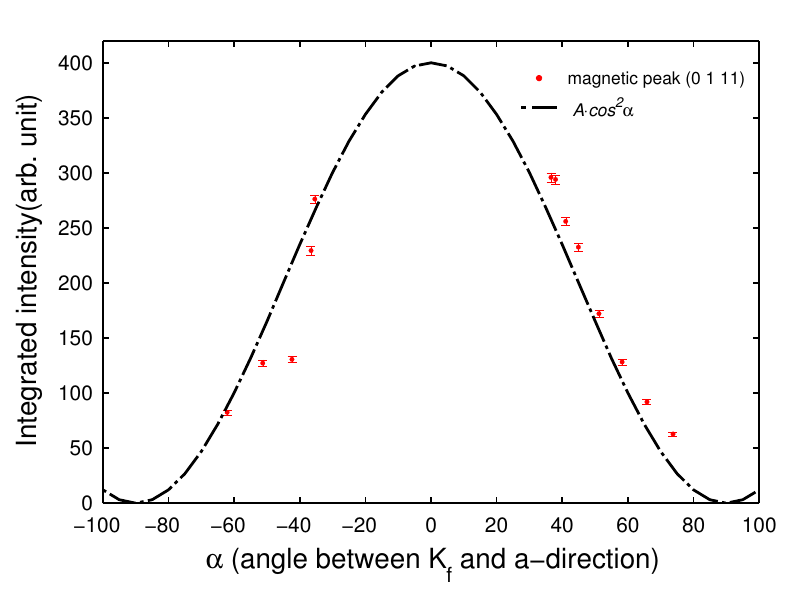}
\caption{The azimuthal dependence of the (0 1 11) magnetic peak intensity. $\alpha$ is the angle between scattered beam and the crystallographic {\it a} direction.}
\label{fig2}
\end{figure}

We also carried out a search for other peaks at various positions, and along high symmetry directions, in reciprocal space. In particular, scans across (0 1 L) with even values of L showed no sign of magnetic peaks, indicating that the two Ir sublattice planes, which are separated by half a unit cell in the {\it c} direction, are antiferromagnetically coupled. Further, no peaks were found between (0 1 11) and (0 0 11), or between (0 1 11) and (1 1 11). These latter observations, together with the determination of the in-plane ordering wavevector to be (0 1), allow us to conclude that the magnetic unit cell size is the same as the structural unit cell. Unfortunately, scans across (0 3 L) and (1 0 L) with L = 11 and 13 were contaminated by nearby strong non-magnetic powder peaks, which prevent us from solving the magnetic structure completely. Nevertheless, the observed peaks do provide strong constraints on the possible magnetic structure of Na$_2$IrO$_3$, as we shall now discuss.

\begin{figure}[ht]
\includegraphics[width=0.48\textwidth]{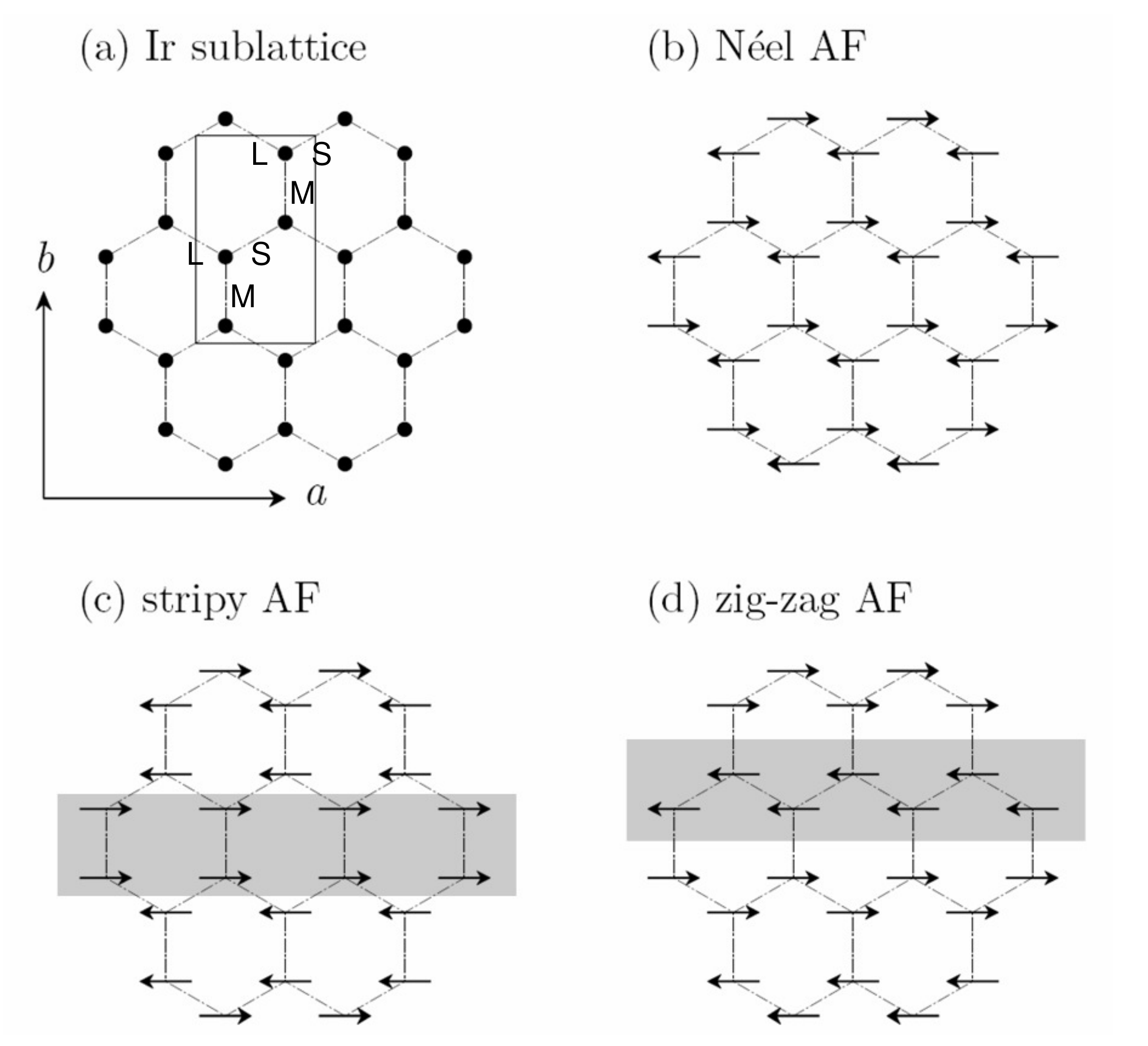}
\caption{Crystal structure and possible antiferromagnetic ordering patterns of the Ir sublattice. (a) crystal structure: the arrows at the left-bottom corner show the in-plane directions of the crystal; the solid-line box indicates the in-plane structural unit cell; different Ir-Ir bonds are labeled with L(long), M(medium) and S(short) with the bond lengths to be {3.30\AA}, {3.23\AA} and {2.86\AA} respectively. (b), (c) and (d): 3 possible magnetic structures. In each case the magnetic unit cell is the same as the structural unit cell shown in (a). The shaded boxes highlight the stripy and zig-zag chain elements accordingly.}
\label{fig3}
\end{figure}

With the determination that the magnetic unit cell is the same as the crystallographic unit cell and that the ordered moment points along the {\it a}-axis, the number of possible AF configurations in Na$_2$IrO$_3$ are reduced to three, shown in Fig. 3. The N\'eel and so-called ``stripy'' states were discussed by Chaloupka {\it et al.}\cite{honeycomb2} and were found by adding an isotropic Heisenberg term to the Kitaev Hamiltonian, while the zig-zag pattern has not been considered in Na$_2$IrO$_3$ before. The non-zero (0 1 L) peaks rule out the N\'eel state, leaving just the stripy and zig-zag patterns as possibilities. 

To further resolve the magnetic ground state, we performed spin density functional theory calculations\cite{compdetails} of four collinear magnetic configurations (FM, N\'eel, stripy and zig-zag) in which the moments are constrained along each of the three crystallographic axes, using the empirical monoclinic C2/{\it c} structure\cite{Yogesh}. The results are summarized in table \ref{tab:tab1}. For all the moment directions, the zig-zag configuration is found to have lower total energy than the stripy configuration. This, in combination with the x-ray analysis, leads us to propose the zig-zag configuration as the most likely ground state of Na$_2$IrO$_3$.

The calculation results also show that for the in-plane directions of the moments ($a$ and $b$), the FM and zig-zag configurations are nearly equally low in energy, whereas the stripy and N\'eel configurations are nearly equally high in energy. The other electronic properties, the band gap, the spin and orbital moments, display the same trend. 
We observe that the FM and zig-zag configurations have ferromagnetically aligned moment in the shortest Ir-Ir bond, while the stripy and N\'eel have the same bond antiferromagnetically aligned (see Fig. 3). We note that the constrained moment directions investigated in this study are not meant to represent the moment direction of the theoretical ground state, which in general could be away from the crystallographic axes and non-collinear.

\begin{table}
\caption{The total energy E$_{\rm{tot}}$ per Ir, the bandgap E$_{\rm{gap}}$ and the absolute value of the spin $| \langle S \rangle |$ and orbital $|\langle L \rangle|$ moment of four collinear magnetic configurations calculated within the LDA+U+SO approximation with the moments constrained along each of the three crystallographic axes ($a$, $b$ and $c$).
}
\begin{ruledtabular}
\begin{tabular}{ccccc}
 config. & E$_{\rm{tot}}$per Ir(meV)  & E$_{\rm{gap}}$(meV) & $|\langle S\rangle|(\mu_B)$ &$|\langle L\rangle|(\mu_B)$ \\
\hline
FM $a$  & 0  & 100  & 0.52  & 0.26 \\
FM $b$ & -17 & 212 & 0.53  & 0.33 \\
FM $c$ & 13  & 0 & 0.52 & 0.25 \\
\hline
N\'eel $a$ & 50  & 196  & 0.33 &  0.31 \\
N\'eel $b$ &  2  & 298  & 0.33 &  0.36 \\
N\'eel $c$ &  9  & 276  & 0.14 & 0.24 \\
\hline
zig-zag  $a$ &  2  & 276  & 0.50  & 0.25 \\
zig-zag  $b$ &  -20  & 324  & 0.51 & 0.32 \\
zig-zag  $c$  & -14  & 261  & 0.51 & 0.29 \\
\hline
stripy $a$ & 42  & 224  & 0.36 & 0.35 \\
stripy $b$ & 2  & 309  & 0.36 & 0.37 \\
stripy $c$ & -2  & 261  & 0.32 & 0.35 \\
\end{tabular}
\end{ruledtabular}
\label{tab:tab1}
\end{table}

We now turn to the temperature dependence of the magnetic ordering, which was monitored by H scans across (0 1 11) peak. The magnetic peak, whose width is almost a constant as function of temperature, was fit to a Lorenztian-squared function. In Fig. 4, we plot the temperature dependence of the integrated intensity obtained from fitting. From the onset of the magnetic peak intensity, $T_N$ is determined to be 13.3($\pm0.1$) K. We note that the order parameter could not be fitted to a simple power law, even for the region near $T_N$. There is a small linear region close to the transition which extrapolates to a lower $T_N = 12.7$ K(see Fig. 4 insert). Above this temperature, a long tail persists up to 13.3 K. One possible explanation for this is a small rounding of T$_N$ arising from sample imperfections, though other explanations are also possible in systems with complex magnetic interactions. For example, similar temperature dependence was observed in the $4f$ heavy femion system CeAs\cite{CeAs}. There the gradual onset of long range AF ordering was attributed to the suppression of the development of long range magnetic ordering due to strong short range spin fluctuations. In Na$_2$IrO$_3$, the presence of short range correlations have also been suggested from magnetic and heat capacity measurements\cite{Yogesh}.

\begin{figure}[ht]
\includegraphics{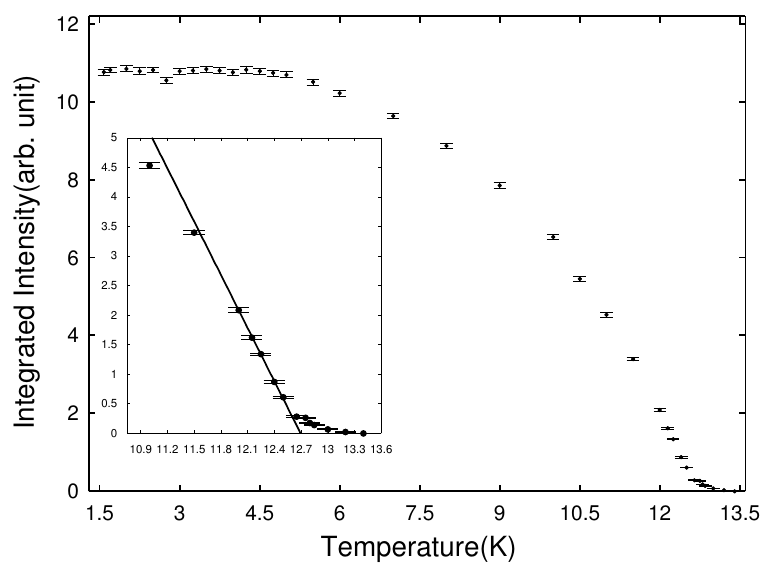}
\caption{Temperature dependence of the integrated intensity of the (0 1 11) peak. Insert: Expanded view of the temperature region near $T_N$.}
\label{fig4}
\end{figure}

We next discuss the significance of these results in the context of the search for an experimental realization of the Kitaev model. There is a general consensus that systems with effective spin 1/2 on a honeycomb lattice with strong SOC, as are present in the Na$_2$IrO$_3$, are good places to look. For example, Chaloupka {\it et al.}\cite{honeycomb2} showed that for such a system the underlying Hamiltonian could be described as a sum of an isotropic Heisenberg term and the Kitaev Hamiltonian. They found a range of possible ground states, from a N\'eel state to the quantum spin liquid. Our data rule out the N\'eel state, implying that the Kitaev term is significant in Na$_2$IrO$_3$ and that it is not governed by a simple Heisenberg Hamiltonian. A second point to emphasize is the observed magnetic ordering is three dimensional, that is the correlation length is as long perpendicular to the planes as within them. The Kitaev model describes a two-dimensional honeycomb lattice and its ground state is expected to be robust to small perturbations. The fact that we are not seeing the spin liquid state suggests that these perturbations, which might be out-of-plane interactions, for example, or finite Heisenberg exchange, are relatively strong in this material.

In conclusion, three-dimensional long-range antiferromagnetic ordering is found in Na$_2$IrO$_3$ by resonant x-ray scattering at Ir $L_3$ edge. The azimuthal dependence of the magnetic scattering intensity indicates that the ordered moment is predominantly along the {\it a}-direction. Two possible ordering patterns, namely stripy and zig-zag, are consistent with the experimental data. Combining the magnetic x-ray analysis with a set of constrained first principles calculations, we propose the ``zig-zag'' phase as the most likely ground state. Given that no real materials have been found to show pure Kitaev physics and that Na$_2$IrO$_3$ was promising from many aspects, understanding our observations from a theoretical perspective could shed light on why the Kitaev model breaks down in Na$_2$IrO$_3$, and thus pave the way towards the final realization of spin-liquid states in real materials.

The work at Brookhaven was supported by the U.S. Department of Energy, Division of Materials Science, under Contract No. DE-AC02-98CH10886. Y.S. would like to thank the Alexander von
Humboldt foundation for support. The work at University of Toronto was supported by the NSERC of Canada.


\begin{thebibliography}{1}
\bibitem{compdetails} We applied the WIEN2K~\cite{Blaha} implementation of the all electron linearized augmented plane wave method in the LDA+U+SO approximation, within the second variational treatment using U=3eV and J=0.6eV. In the three antiferromagnetic configurations, the moments connected by half a unit cell in the $c$ direction, were aligned antiferromagnetically. The basis size was determined by RKmax=8.0 and the Brillouin zone was sampled with a 5x5x2-mesh.
\end{thebibliography}
\end{document}